# Multi-Objective Personalized Product Retrieval in Taobao Search

Yukun Zheng, Jiang Bian, Guanghao Meng, Chao Zhang, Honggang Wang, Zhixuan Zhang, Sen Li,
Tao Zhuang, Qingwen Liu, and Xiaoyi Zeng
Alibaba Group, Beijing, China
zhengyk13@gmail.com

## ABSTRACT

In large-scale e-commerce platforms like Taobao, it is a big challenge to retrieve products that satisfy users from billions of candidates. Recently, plenty of works in this domain have achieved significant improvements by enhancing embedding-based retrieval (EBR) methods, including the Multi-Grained Deep Semantic Product Retrieval (MGDSPR) model [16] in Taobao search engine. However, we find that MGDSPR still has problems of poor relevance and weak personalization compared to other retrieval methods in our online system, such as lexical matching and collaborative filtering. These problems promote us to further strengthen the capabilities of our EBR model in both relevance estimation and personalized retrieval. Existing EBR models learn to rank the positive item before negatives in each *single-positive* training sample and do not take the relations between multiple positive and negative items in the same page view into account. This damages the retrieval performance of EBR models. In this paper, we propose a novel **M**ulti-**O**bjective **P**ersonalized **P**roduct **R**etrieval (MOPPR) model with four hierarchical optimization objectives: relevance, exposure, click and purchase. We construct *entire-space multi-positive samples* to train MOPPR and adopt a modified softmax loss for optimizing multiple objectives. Results of extensive offline and online experiments show that MOPPR outperforms baseline methods on evaluation metrics of relevance estimation and personalized retrieval. MOPPR achieves 0.96% transaction and 1.29% GMV improvements in a 28-day online A/B test. Since the Double-11 shopping festival of 2021, MOPPR has been fully deployed in mobile Taobao search, replacing the previous MGDSPR. Finally, we discuss several advanced topics of our deeper explorations on multi-objective retrieval and ranking to contribute to the community.

## 1 INTRODUCTION

The core goal of search engines in e-commerce platforms is to provide products that can satisfy users. With the development of Taobao platform in recent years, the number of products on Taobao has grown rapidly to billions. This brings great challenges to the search engine, especially to our retrieval system. Benefitted from the development of deep learning, a number of embedding-based retrieval (EBR) methods [10, 16, 21, 32] have been proposed to address this problem and achieved great success in practical e-commerce platforms such as Amazon [21] and JD [32]. We also built our two-tower embedding-based retrieval method named MGDSPR [16] in Taobao search system. By retrieving personalized products with good relevance, MGDSPR achieved significant transaction improvement for Taobao. Figure 1 shows the architecture of the current Taobao search engine with two sub-systems: multi-channel retrieval and cascade ranking. When a user submits a query, the multi-channel retrieval system, including our EBR model, retrieves a collection of candidate products. Then the cascade ranking system screens and sorts the retrieved products step by step, and finally presents a page of products to the user. A strict relevance control module is deployed in the cascade ranking system to ensure the relevance of search results with respect to the query. During a long-time online experiment, we find two problems of MGDSPR:

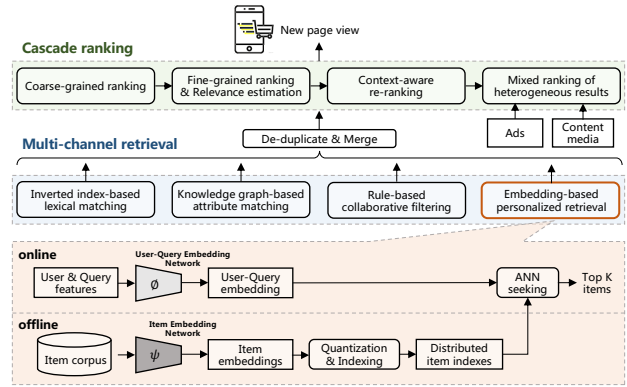

**Figure 1: The architecture of Taobao search engine with our embedding-based retrieval method.**

- **Poor relevance**: Based on massive manual relevance annotations, we find that products retrieved by MGDSPR have lower relevance than products of invert index-based lexical matching [23, 37] on average.
- **Weak personalization**: We find the proportions of products retrieved by MGDSPR in the sets of exposed, clicked and purchased items are smaller compared to lexical matching and collaborative filtering [18]. This indicates the limitation of MGDSPR on retrieving personalized products that satisfy users.

To address the aforementioned problems, we introduce multiple objective optimization into EBR models to retrieve relevant items that meet the personalized needs of users as much as possible. According to the findings of our data analysis, we assume that only relevant products can be exposed, further clicked and finally purchased. When a user submits a query in the search engine, given the relevance probability $P(Rel = 1)$ and the conditional exposure probability $P(Exp = 1|Rel = 1)$ of an item (product) $i$, the probability of item $i$ exposing to the user is $P(Exp = 1) = P(Exp = 1|Rel = 1)P(Rel = 1)$. Next, given its click probability $P(Click = 1|Exp = 1)$ and conversion probability $P(Purchase = 1|Click = 1)$, its purchase probability is finally defined as:

$$\begin{aligned} P(Purchase = 1) = &P(Purchase = 1|Click = 1) \\ &P(Click = 1|Exp = 1) \\ &P(Exp = 1|Rel = 1)P(Rel = 1) \end{aligned} \quad (1)$$

Therefore, to retrieve items that satisfy users' purchase needs, the EBR model needs to retrieve products with good relevance as well as high exposure, click and conversion probabilities.

However, two challenges on sample construction and loss function hinder us from training an effective EBR model with the multiple objectives above. A training sample of existing EBR models [10, 16, 32] consists of a single positive item (e.g., a clicked or purchased item) and a number of negative items, which we call a **single-positive** sample. The loss functions of these models are optimized to maximize the score of the positive item and minimize the scores of the negative items. Data fusion with sample weighting [10, 16] is a popular method to achieve multi-objective optimization for these models. We consider that this method is not optimal because it does not let models learn the relations between different kinds of positives and negatives at the same time. It is necessary to design a new sample construction method and an applicable loss function for the multi-objective retrieval model.

In this paper, we propose a novel **M**ulti-**O**bjective **P**ersonalized **P**roduct **R**etrieval (MOPPR) model with four optimization objectives: relevance, exposure, click and purchase. We construct new **multi-objective and multi-positive** samples to train MOPPR. Each sample is generated from a page view in our search engine and includes all impression items in this page view, several sampled under-impressions and a number of random negatives. We adopt a modified softmax loss function for the multi-objective optimization. We conduct extensive offline and online experiments to compare MOPPR with our online baseline MGDSPR and a classical baseline $\alpha$-DNN. Experimental results show that MOPPR significantly outperforms baselines on all offline metrics and online performances. We have fully deployed MOPPR in mobile Taobao Search to serve hundreds of millions of users since the Double-11 shopping festival of 2021, better satisfying users' search demands and bringing significant GMV improvements. We further discuss our deeper explorations on hard mining, GMV maximization retrieval, and the cascade model of multi-objective retrieval and ranking.

## 2 RELATED WORKS
### 2.1 Deep Matching and Retrieval

With the rapid development of deep learning, a large number of neural network-based models have been proposed to solve the semantic gap caused by traditional information retrieval (IR) methods [20]. DSSM [11] first built neural networks with two towers to generate semantic representations of queries and documents. The following works include CDSSM [25], DRMM [9], and CLSM [24]. These works incorporated traditional IR matching techniques (e.g., sliding window, query word importance, exact matching) into the neural networks to improve two-tower embedding-based retrieval models. How to retrieve more personalized and semantically relevant results remains two major challenges in modern search matching works based on deep learning.

Recently, embedding-based retrieval (EBR) technologies have been widely adopted in the modern recommendation, search, and advertising systems [4, 15, 16, 21, 21, 35]. For social networks, Facebook proposed an embedding modeling framework for personalized semantic search and an online retrieval system based on embedded expressions [10]. For video retrieval, Google corrected negative sampling to solve the sampling bias of batch softmax [31]. For e-commerce retrieval, Amazon has developed a two-tower model to fix the semantic gap problem in the semantic product retrieval matching engine [21]. Alibaba proposed the multi-grained deep semantic product retrieval [16] to solve the low relevance issue caused by the EBR method. However, most of the existing EBR models only utilize a single clicked or purchased item as the optimization target in each training sample, thus it is not optimal to learn the relations of all kinds of items and is difficult to retrieve more personalized and relevant items.

### 2.2 Multi-Objective Optimization

In the field of machine learning, multi-objective optimization has been developed for many years and obtained impressive performance. The existing multi-objective optimization can be categorized into scalarization [28] and evolutionary algorithms [36]. Scalarization depends on manually assigned weights for multi-objective learning. Evolutionary algorithms transform the multi-objective optimization into a problem of finding the Pareto efficient solution [22]. In recent years, multi-objective optimization has attracted much attention in the modern recommendation and search systems. A lot of related studies [2, 14, 27] focus on how to model the relations between the multiple objectives (e.g., relevance, diversity, purchase likelihood). Long et al. [19] combined relevance scores and purchase prediction scores to perform search rankings. Svore et al. [26] optimized multiple goals based on the relevance judgment and click feedback of human annotators. Dai et al. [5] generated a mixed label to incorporate the relevance and freshness of items for ranking optimization. Karmaker Santu et al. [13] utilized feedback signals (such as click-through rate, order rate, revenue and adding to the shopping cart) as the training and evaluation criteria. Carmel et al. [3] applied the label aggregation method to aggregate labels of training samples with different targets into a single label, reducing the problem to single-objective optimization. Lin et al. [17] proposed a Pareto-Efficient algorithmic framework for the e-commerce recommendation task to optimize multiple objectives. Xie et al. [29] proposed a novel Personalized Approximate Pareto-Efficient Recommendation (PAPERec) framework for multi-objective recommendation to capture users' objective-level preferences and enhance personalization. However, most multi-objective optimization methods focused on the ranking problem of search or recommendation and have not been widely used in product matching and retrieval. In this work, we explore multi-objective retrieval optimization, which takes into account both the personalization and semantic relevance of products.

## 3 PROBLEM FORMULATION

In this section, we first formulate the personalized product retrieval problem and the mathematical notations used in the rest of this paper. Let $\mathcal{U} = \{u_1, \cdots, u_{|\mathcal{U}|}\}$ denotes the set of users, $Q = \{q_1, \cdots, q_{|Q|}\}$ denotes the set of search queries, and $\mathcal{I} = \{i_1, \cdots, i_{|\mathcal{I}|}\}$ denotes the set of items (products), where $|\mathcal{U}|$, $|Q|$, $|\mathcal{I}|$ are the numbers of distinct users, queries and items, respectively. We segment query $q$ into short terms: $q = \{w_1, \cdots, w_{|q|}\}$, where $w_i$ is the $i$-th term of $q$ and $|q|$ is the number of terms. Let $\mathcal{B} = \{i_1^u, \cdots, i_{|\mathcal{B}|}^u\}$ denotes the historical behaviors of the user $u$, including $u$'s clicked, collected, and purchased items as well as items added to the shopping cart. Further, we divide the user's behaviors into three non-overlapping collections according to the time interval from the current time: real-time behaviors $\mathcal{B}_r$ in the past one day, short-term behaviors $\mathcal{B}_s$ from the past second day to the past tenth day, and long-term behaviors $\mathcal{B}_l$ from the past eleventh day to the past one month, i.e., $\mathcal{B} = \mathcal{B}_r \bigcup \mathcal{B}_s \bigcup \mathcal{B}_l$.

Our goal for the personalized product retrieval problem is to return the items that are most likely to satisfy the user $u$ after $u$ submits a query $q$ to the search engine. In practical systems, retrieval models usually select the top $K$ items from the item corpus $\mathcal{I}$ according to the prediction scores. Formally, given a quadruple $\langle u, q, \mathcal{B}, i \rangle$, the retrieval model predicts the score $z$ as follows:

$$z = \mathcal{F}(\phi(u, q, \mathcal{B}), \psi(i)) \quad (2)$$

where $\mathcal{F}(\cdot)$ is the score function, $\phi(\cdot)$ and $\psi(\cdot)$ are the user and item embedding functions, respectively. In this paper, we follow the two-tower retrieval framework [10] and adopt the inner product operation as $\mathcal{F}(\cdot)$ performed between the user-query and item embedding vectors predicted by $\phi(\cdot)$ and $\psi(\cdot)$.

## 4 MODEL

Figure 2 shows the architecture of MOPPR. MOPPR generates the user-query embedding and the item embedding independently and finally performs the inner product between them as the prediction score. Different from the existing EBR models trained on single-positive samples, MOPPR learns on multi-positive samples with a multi-objective loss function.

### 4.1 User-Query Tower

The user-query tower consists of three units: query semantic unit, user behavior attention unit, and embedding prediction unit.

*4.1.1 Query Semantic Unit.* With the embedding matrix of the query term sequence $e^q = \{e^{w_1}, \cdots, e^{w_{|q|}}\} \in \mathbb{R}^{|q| \times d}$ as the input, the query semantic unit generates three kinds of query representations: mean-pooling representation $Q_m \in \mathbb{R}^{1 \times d}$, self-attention representation $Q_s \in \mathbb{R}^{1 \times d}$ and personalized representation $Q_p \in \mathbb{R}^{1 \times d}$. The overall query representation $Q_o \in \mathbb{R}^{1 \times 3d}$ is obtained as follows:

$$Q_o = concat(Q_m, Q_s, Q_p) \quad (3)$$
$$Q_m = mean\_pooling(e^q) \quad (4)$$
$$Q_s = max\_pooling(self\_atten(e^q))$$
$$= max\_pooling(softmax(\frac{e^q \cdot (e^q)^T}{\sqrt{d}}) \cdot e^q) \quad (5)$$
$$Q_p = softmax(\frac{(e^u W_1 + b_1) \cdot (e^q)^T}{\sqrt{d}}) \cdot e^q \quad (6)$$

where $mean\_pooling$, $max\_pooling$ and $concat$ are the average, maximum and concatenation operation, and $e^u$ is the user representation obtained by concatenating all embeddings of the user profile features, $W_1 \in \mathbb{R}^{d \times d'}$ and $b_1 \in \mathbb{R}^{1 \times d'}$ are the parameters of a fully connected layer for $e^u$. With this query semantic unit, we fuse generic and personalized representations of the query together, which helps capture rich semantic information in the query.

*4.1.2 User Behavior Attention Unit.* Users' historical behaviors reflect their rich interests. However, there are also many noisy behaviors that harm the model performance. To remove the impact of these noisy behaviors, we first train a deep learning model to predict the relevant categories of queries, and then filter out the historical behaviors that do not belong to the predicted relevant categories of the query. This operation is called *Category Filtering* [12]. Next, we adopt the query attention mechanism [16] to effectively mine useful information from the reserved historical behaviors. We do not use target-item attention [7, 34] because it is inapplicable in the two-tower retrieval framework where item features cannot interact with user historical behaviors. We use the embeddings of user profile features and query features as the *Query* of the attention unit and the user's historical behavior sequence as the *Key* and *Value*. We perform the same query attention on three collections of user behaviors $\mathcal{B}_r$, $\mathcal{B}_s$ and $\mathcal{B}_l$. Taking $\mathcal{B}_r$ for example, each item $i \in \mathbb{R}^{1 \times d_i}$ in $\mathcal{B}_r$ is represented by concatenating its ID and side information embeddings together along the last axis. Next, the embedding matrix of the real-time behavior sequence, denoted as $\mathbb{B}_r = (i_1, \cdots, i_{|\mathcal{B}_r|}) \in \mathbb{R}^{|\mathcal{B}_r| \times d_i}$, is aggregated by the query attention. The output $H_r$ of the query attention is defined as follows:

$$H_r = softmax(\frac{Q \cdot (\mathbb{B}_r)^T}{\sqrt{d_i}}) \cdot \mathbb{B}_r \quad (7)$$
$$Q = concat(Q_o, e^q, e^u) W_r + b_r \quad (8)$$

where $Q_o$, $e^q$, $e^u$ are the query semantic representation generated from the query semantic unit, the concatenated embeddings of query side information (such as query frequency and relevant categories), and the concatenated embeddings of the user profile features, respectively. $W_r \in \mathbb{R}^{d_{uq} \times d_i}$ and $b_r \in \mathbb{R}^{1 \times d_i}$ are the parameters of a fully connected layer for the real-time behavior sequence.

Next, we concatenate the representations of real-time, short-term and long-term behaviors, i.e., $H_r$, $H_s$ and $H_l$, as the final representation of user historical behaviors $H_B$:

$$H_B = concat(H_r, H_s, H_l) \quad (9)$$

*4.1.3 Embedding Prediction Unit.* After processing all query and user features, we concatenate all representations of user and query together and feed the new representation $e^{uq}$ into a Multilayer Perceptron ($MLP_{uq}$) with four layers. Each layer of the first three layers consists of a fully connected layer ($FC$), a layer normalization layer ($LN$) and a Leaky ReLU layer ($LReLU$). The last layer of MLP is a fully connected layer ($FC$). The user-query representation $H_{uq}$ is obtained by normalizing the output vector of MLP to a unit vector:

$$H_{uq} = l2\_norm(MLP_{uq}(e^{uq})) \qquad (10)$$
$$e^{uq} = concat(e^u, e^q, Q_o, H_B) \qquad (11)$$
$$MLP_{uq} = \{(FC, LN, LReLU) \times 3, FC\} \qquad (12)$$

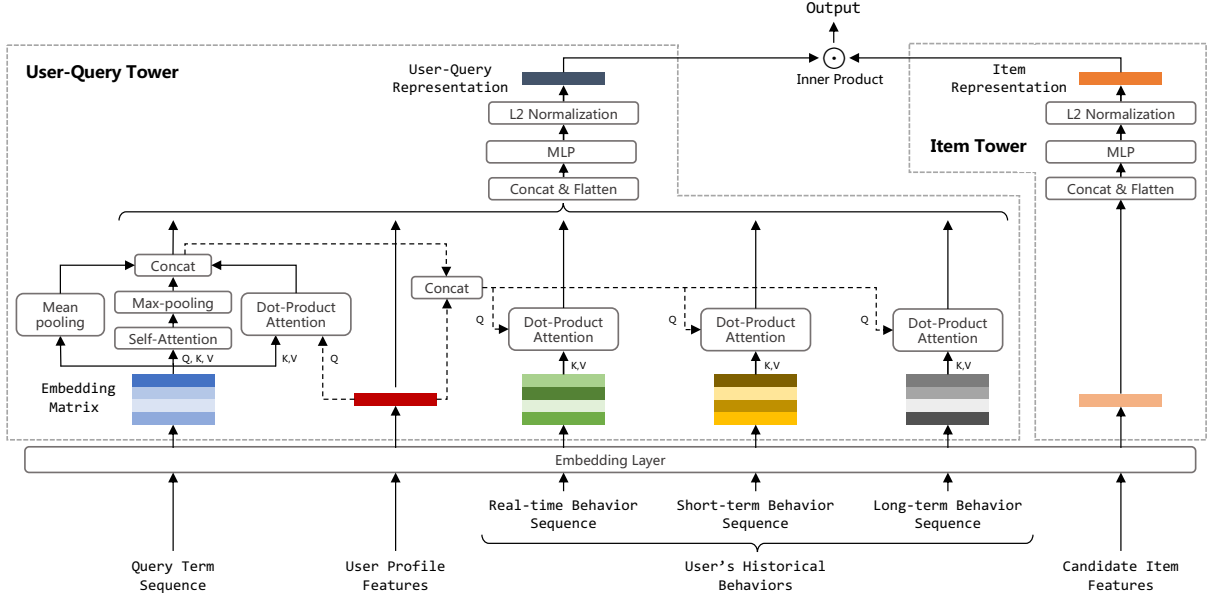

Figure 2: The architecture of Multi-Objective Personalized Product Retrieval model (MOPPR).

## 4.2 Item Tower

In the item tower, we first get the item embedding $e^i$ by concatenating all embeddings of item ID and item side information together. For the item title feature, we utilize the mean-pooling operation on embeddings of title terms to get its embedding. Then we feed the item embedding $e^i$ into another Multilayer Perceptron ($MLP_i$) with four layers. Thus, the final item representation $H_i$ is defined as:

$$H_i = l2\_norm(MLP_i(e^i)) \qquad (13)$$
$$MLP_i = \{(FC, LN, LReLU) \times 3, FC\} \qquad (14)$$

## 4.3 Multi-Objective Optimization

To enable MOPPR to retrieve products that satisfy multiple objectives, we improve its sample construction and propose a new multi-objective loss function. Figure 3 shows the differences between the existing retrieval models and our multi-objective retrieval model in sample construction and optimization objectives.

*4.3.1 Sample Construction.* By learning from massive behavior log data, the retrieval model fits users' click and purchase behaviors and models users' personalized preferences. As Figure 3 shows, a training sample for traditional EBR models [16] consists of features of the user, query, a positive item and a number of negative items. The positive item in each sample is usually the user's clicked or purchased item. The negative items are usually randomly sampled from the item corpus and shared within the same mini-batch. The optimization goal of traditional EBR models is to maximize the prediction score of the positive item. Therefore, we call these kinds of models *single-positive models*.

In contrast, MOPPR is trained on page-level training samples with multiple objectives and multiple positive items. One sample of MOPPR contains all items presented in a page view of Taobao search engine. To be specific, in each training sample of MOPPR, there are features of the user, query, and several kinds of items:

- **Impressions**: $N$ items presented in the page view, including the items clicked or purchased by the user. These presented items were sorted at the top of all candidates by the cascade ranking system and are predicted to have the largest probabilities to satisfy the user.
- **Under-Impressions**: $M$ items that were retrieved by the online multi-channel retrieval system but were sorted after the impressions and were not presented to the user. We sample $M$ items from thousands of under-impressions per page view through an online logging system. We only reserve the under-impressions whose ranking positions are greater than 150 in the cascade ranking stage.
- **Random negatives**: $L$ Items that are randomly sampled from the item corpus as negatives. We adopt the in-batch negative sharing method [16]. Each sample takes the random negatives of the other samples within the same mini-batch as its negatives. So, given the batch size $B$, we only need to sample $\frac{L}{B}$ random

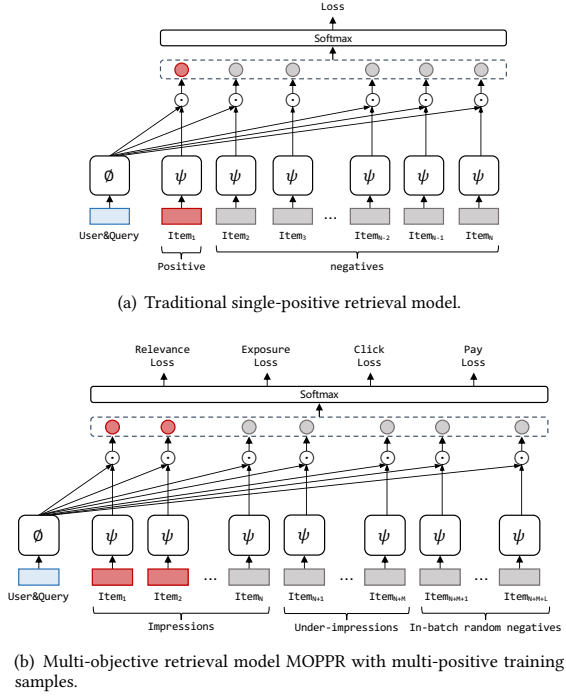

(a) Traditional single-positive retrieval model.

(b) Multi-objective retrieval model MOPPR with multi-positive training samples.

**Figure 3: Differences between the single-positive and multi-objective retrieval models in sample construction and optimization objectives. Items in red denotes the clicked or purchased items.**

negatives for each training sample, which greatly reduces the computing cost and storage resources.

For all items in each sample, we adopt four kinds of binary labels corresponding to four optimization objectives of MOPPR:

- **Relevance Label**: This label denotes whether the item is relevant to the query. Our online relevance estimation model predicts the relevance scores for all impressions and under-impressions, which have 0.915 AUC with human relevance judgments [30]. We obtain the relevance scores of impressions and under-impressions from the log data and discretize them into binary labels. The relevance labels of random negatives are set to 0.
- **Exposure Label**: This label denotes whether the item is presented in the page view, which is determined by the cascade ranking system. The labels are 1 for impressions and 0 for all other items.
- **Click Label**: This label denotes whether the item is clicked by the user in this page view.
- **Purchase Label**: This label denotes whether the item is purchased by the user in this page view.

*4.3.2 Multi-Objective Loss Function.* For the loss function of MOPPR, we follow Li et al. [16] and adopt softmax function, which compares all items of a training sample in a holistic view. The softmax function also shows better performance and faster convergence speed than pairwise functions such as hinge loss used by Huang et al. [10]. Formally, our MOPPR model has four optimization objectives $O = \{Relevance, Exposure, Click, Purchase\}$. For any objective $o \in O$, given a user $u$ with his/her historical behaviors $\mathcal{B}$ and a query $q$, candidate items $I = \{i_1, i_2, \cdots, i_{N+M+L}\}$ have labels $y^o = \{y_1^o, y_2^o, \cdots, y_{N+M+L}^o\}$. The number of positive items in $y^o$ is denoted as $|o^+|$. The loss for objective $o$ is calculated as follows:

$$\hat{y}(i_k|u, q, \mathcal{B}) = \frac{\exp(\mathcal{F}(\phi(u, q, \mathcal{B}), \psi(i_k))/\tau)}{\sum_{i_j \in I} \exp(\mathcal{F}(\phi(u, q, \mathcal{B}), \psi(i_j))/\tau)} \quad (15)$$

$$\hat{y}_k = \min(\hat{y}(i_k|u, q, \mathcal{B}) \cdot |o^+|, 1) \quad (16)$$

$$L_o(\nabla) = \begin{cases} -\sum_{i_k \in I} y_k \log(\hat{y}_k), & when \ |o^+| > 0 \\ 0, & when \ |o^+| = 0 \end{cases} \quad (17)$$

where $\tau$ is the temperature parameter. By multiplying the softmax probability $\hat{y}(i_k|u, q, B)$ by the positive number $|o^+|$ and limiting its maximum value to 1 in Equation 16, the minimum value of loss $L_o(\nabla)$ becomes zero. Another advantage of this operation is to help the model stop optimizing those positive items with excessively large scores. For example, given a sample with two click positives, if one click positive has a much larger prediction score than all the other items including the other click positive, its probability after softmax in Equation 15 is more likely to be greater than 1 and will be limited to 1 in Equation 16. Therefore, there will be no backward gradient for optimizing this click positive item. In other words, this operation helps the model to balance the relations among multiple positive items, which can not be considered in single-positive retrieval models. The total loss $L(\nabla)$ of MOPPR is the weighted sum of four objectives' losses, defined as follows:

$$L(\nabla) = \sum_{o \in O} w_o L_o(\nabla) \quad (18)$$

where $w_o$ is the loss weight for the objective $o$. These loss weights help adjust the impact of different optimization objectives. Considering that the positive numbers of four objectives are quite different and gradually reduced in the order of relevance, exposure, click and purchase objectives, it's important to adjust these loss weights to overcome the impact of this bias.

### 4.4 Discussion

Here we explain the main differences between our MOPPR model and existing retrieval models, and discuss the advantages of MOPPR over them:

- **Entire-space training samples**: Different from existing EBR models trained on single-positive samples, MOPPR learns users' search preferences on multi-objective and multi-positive samples. This boosts retrieval performance by providing more hard negatives. Besides, this helps MOPPR to learn the relations between multiple positive and negative items at the same time, such as multiple clicks or impressions occurred in the same page view.
- **Hierarchical retrieval objectives**: For single-positive EBR models, data fusion and sample weighting are usually adopted to achieve multi-objective optimization. MOPPR optimizes multiple objectives more straightforwardly. Hierarchical objectives help MOPPR maximize the scores of different kinds of positive items and learn to sort items in the following optimal order: purchased items > clicked but non-purchased items > non-clicked impressions > relevant under-impressions > irrelevant items.

# 5 OFFLINE EXPERIMENT

## 5.1 Dataset

We use online click and purchase log data collected from the search engine of *Mobile Taobao APP* to construct our training data regularly every day. We train retrieval models only on the samples with more than one click in the page view. All input features is divided into four types: ID, textual, attribute and statistical features. ID features represent IDs of the user, query and item. Textual features are represented as sequences of short terms, such as query text and item title. Attribute features include the user' profile information, the relevant categories of queries and the side information of items such as brand, category and seller. Statistical features denote the user-item interaction statistics on the user/query/item level. Statistical features were proven to play an important role in the CTR prediction of the ranking stage [8, 33]. Therefore, we also take them as features in MOPPR.

## 5.2 Baselines

In this experiment, we adopt the previous EBR model MGDSPR [16] in Taobao search engine and $\alpha$-DNN [4] as baselines. The two baselines are trained on singe-positive samples where the positive items are users' clicked items. Thus, we call them **single-objective baselines**. To fairly compare them with MOPPR, we implement them to take the same input features as MOPPR. MGDSPR is one of the strongest baselines in industrial product retrieval within our knowledge. It has stronger relevance capability than its predecessor model in Taobao search by smoothing noisy training data, mining hard relevance negatives and using a multi-granular semantic network. $\alpha$-DNN is a classical EBR model which was proven to be effective in many retrieval tasks. The parameters of MOPPR and two baselines have been incrementally trained for over 2 years. Training on so much data is extremely time-consuming and these models are well-fitted to our online dataset. Considering that it is unfair to compare other cold-start baselines with them, we do not adopt other baseline methods.

## 5.3 Experimental Setup

We implement retrieval models using TensorFlow [1] with Adagrad optimizer [6]. The models were trained on the distributed TensorFlow platform using 80 parameter servers and 1000 workers with 20 CPUs per worker. The batch size is set to 1024 and the initial learning rate is 0.01. The dimension of user-query and item representation $H_{uq}$ and $H_i$ is set to 128. The temperature $\tau$ is set to 0.02 because of the best performance in the grid search. Given the positive item number $|o^+|$ of an objective $o \in O$ within a mini-batch, we set its loss weight $w_o$ to $\frac{1}{|o^+|}$ to balance the effect of four objectives. The numbers of impressions, under-impressions and random negatives per training sample are 10, 10 and 20,480 (i.e., 20 × 1024).

## 5.4 Evaluation

To evaluate the performance of retrieval models, we first construct the evaluation dataset sampled from the log data with three parts: (1) 0.5 million records from search click logs; (2) 0.5 million records from search purchase logs; (3) 0.5 million records whose queries are relevant to the items purchased later by the user in somewhere else such as purchased in the recommender system. One record denotes a search request for a new page view. The corpus of candidate items for retrieval models contains about 0.1 billion active items in Taobao platform.

We evaluate the retrieval performance of models according to these metrics: (1) $recall@K$: whether the model retrieves the target (clicked and purchased) items within the top $K$ retrieval set; (2) $nDCG@K$: whether the model sorts the target items at the top of the retrieval set; (3) Good rate $P_{good}$: the proportion of items with good relevance in the top $K$ retrieval set. The first two metrics are to evaluate the capability of models in personalized retrieval and the last metric is to examine the capability of models in relevance estimation. Formally, given the user's click or purchased items as target items $T = \{t_1, \cdots, t_N\}$ and the top $K$ retrieval set $I = \{i_1, \cdots, i_K\}$ predicted by the model, $recall@K$, $nDCG@K$ and $P_{good}$ are defined as follows:

$$recall@K = \frac{\sum_{i \in I} \mathcal{I}(i \in T)}{N} \quad (19)$$

$$nDCG@K = \frac{\sum_{k=1}^{K} \frac{\mathcal{I}(i_k \in T)}{log_2(k+1)}}{\sum_{k=1}^{N} \frac{1}{log_2(k+1)}} \quad (20)$$

$$P_{good} = \frac{\sum_{i \in I} \mathcal{R}(i)}{K} \quad (21)$$

where $\mathcal{I}(i \in T)$ returns 1 when $i$ is in the target set $T$, otherwise it returns 0, and $\mathcal{R}(\cdot)$ is a relevance indicator that judges whether the item is of good relevance or not. We use our online well-trained relevance model as the relevance indicator to rate the item's relevance. We set $K$ to 6,000 as same as the retrieval number in our online system. Because purchases bring revenues to the platform, we pay special attention to the metrics $recall@K$ and $nDCG@K$ for items purchased in search, denoted as $recall_p@K$ and $nDCG_p@K$. The two metrics are calculated on 0.5 million evaluation records from search purchase logs.

## 5.5 Experimental Results

*5.5.1 Comparison with Baselines.* We compare MOPPR with baselines and report the results in Table 2. We can see that all evaluation metrics of MOPPR are better than those of MGDSPR and $\alpha$-DNN. And the t-test results show that these improvements are significant. It is worth pointing out that MOPPR achieves larger improvements in $recall_p@K$ and $nDCG_p@k$ over MGDSPR than improvements in $recall@K$ and $nDCG@k$, which demonstrates its stronger capability in personalized retrieval. The 5.7% $P_{good}$ improvement over MGDSPR reveals that MOPPR also is better at relevance estimation than our online baseline MGDSPR.

*5.5.2 Effect of Multi-Objective Losses.* We conduct an ablation study to investigate the effect of multi-objective losses on the retrieval performance of MOPPR. We remove each of the losses for the four optimization objectives separately to train MOPPR and examine its performances. The experimental results are reported in Table 1. Although MOPPR without relevance loss achieves slightly better performance in *recall* and *nDCG* metrics, it gets poor result in $P_{good}$ with a 5.2% drop compared to the original MOPPR. Besides, when we remove the exposure/click/purchase loss from MOPPR, results of evaluation metrics show that model performance suffers to varying degrees. The exposure loss helps the retrieval model learn the ranking patterns of the downstream cascade ranking system and our results show its effectiveness. In summary, the experimental results prove that the combination of our four losses is reasonable and effective in improving the performance of retrieval models.

Table 1: Model performance in the ablation study, where $K$ is 6,000.

| Model | $recall@K$ | $nDCG@K$ | $recall_p@K$ | $nDCG_p@K$ | $P_{good}$ |
| --- | --- | --- | --- | --- | --- |
| MOPPR | 0.936 | 0.254 | 0.977 | 0.334 | **0.456** |
| without relevance loss | 0.938(+0.2%) | **0.265(+1.2%)** | 0.977(+0.1%) | **0.346(+1.2%)** | 0.404(-5.2%) |
| without exposure loss | 0.921(-1.5%) | 0.252(-0.2%) | 0.961(-1.6%) | 0.333(-0.1%) | 0.455(-0.1%) |
| without click loss | 0.921(-1.5%) | 0.234(-1.9%) | 0.967(-1.0%) | 0.310(-2.4%) | 0.450(-0.6%) |
| without purchase loss | 0.924(-1.2%) | 0.217(-3.6%) | 0.965(-1.2%) | 0.271(-6.2%) | 0.455(-0.1%) |
| without non-clicked impressions (NCI) | **0.940(+0.4%)** | 0.259(+0.5%) | 0.977(-0.1%) | 0.332(-0.2%) | 0.449(-0.7%) |
| without under-impressions (UI) | 0.925(-1.1%) | 0.235(-1.9%) | 0.968(-0.9%) | 0.306(-2.7%) | 0.423(-3.3%) |
| without NCI and UI | 0.936(-0.4%) | 0.223(-3.1%) | 0.976(-0.3%) | 0.273(-6.1%) | 0.424(-3.2%) |

Table 2: Comparison with baselines, where $K$ is 6,000. All improvements are statistically significant examined by paired t-test with $p < 0.01$.

| Model | $recall@K$ | $nDCG@K$ | $recall_p@K$ | $nDCG_p@K$ | $P_{good}$ |
| --- | --- | --- | --- | --- | --- |
| $\alpha$-DNN | 0.858 | 0.188 | 0.920 | 0.230 | 0.327 |
| MGDSPR | 0.929 | 0.230 | 0.967 | 0.286 | 0.399 |
| MOPPR | **0.936** | **0.254** | **0.977** | **0.334** | **0.456** |

5.5.3 *Effect of Sample Construction.* We investigate the effect of our sample construction. We separately remove the non-clicked impressions (NCI), under-impressions (UI), and both NCI and UI from the training samples and train MOPPR. As Table 1 shows, when we remove NCI, the metrics $recall@K$ and $nDCG@K$ of MOPPR gets better, while $recall_p@K$ and $nDCG_p@K$ do not get improved and the relevance metric $P_{good}$ becomes slightly worse than the original MOPPR. It is difficult to judge the effect of NCI based on the offline experimental results alone. However, when we publish it to the online system, we find that *MOPPR without NCI* performs strikingly worser than the original MOPPR in the online A/B test with more than 0.5% clicks and transactions drop. The items retrieved by *MOPPR without NCI* have lower click and conversion probabilities when presented to users compared to the original MOPPR on average. We consider that the non-clicked impressions play a role as hard negatives in the click objective and are regarded as positives in the exposure objective. This helps the model have a stronger capability to retrieve items that attract clicks and purchases from users. When we remove UI or remove NCI and UI together from the training data, results of all metrics become significantly worse, indicating that adding under-impressions into the training samples boosts the model's performance.

5.5.4 *Hyper-parameter Analysis.* We conduct a hyper-parameter analysis on the number of in-batch shared random negatives $L$. We try different values of $L$ from 2,048 to 102,400. All metrics perform in the same pattern when $L$ varies, so we only report the results of $recall@K$ and $P_{good}$ here. As shown in Figure 4, MOPPR performs the best in $P_{good}$ and $recall@K$ when $L$ is set to 20,480 and 40,960, respectively. The model performances get harmed when the negative number $L$ is too small or too large. If $L$ is too small, the model cannot learn accurate item representations over the entire collection of items. If $L$ is too large, we find that more relevant items that users have interacted with in the past are included as negatives in our dataset with a 0.04% chance. These items are kind of noise in data and damage the model performance.

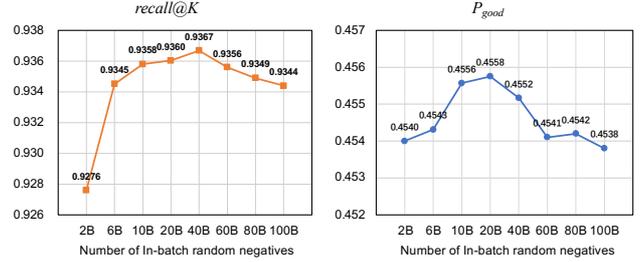

Figure 4: The effect of the number of in-batch shared random negatives, where the batch size $B$ is set to 1,024.

## 6 ONLINE A/B TEST

As shown in Figure 1, with a well-trained retrieval model, we predict the offline embeddings of all candidate items by using its item embedding network. These item embeddings are quantized from FLOAT32 to INT8 and indexed in a hierarchical clustering index tree. When a user searches in our system, the online user-query embedding network predicts the user-query embeddings in real time. Then the approximate near neighbor (ANN) system will seek and return a collection of $K$ items nearest the given user-query embedding. $K$ is set to 6,000 in our system.

We deploy MOPPR online and examine its performance compared to MGDSPR. We conduct a fair online A/B test in mobile Taobao search for 28 days during September 2021. We examine these online metrics: (1) #Transactions: the number of transactions; (2) GMV: the transaction amount; (3) $P_{good}$ and $P_{h\_good}$: the proportion of items with good relevance in the set of exposed items, rated by the relevance model and human annotators, respectively; (4) $P_{exposure}$, $P_{click}$, $P_{purchase}$ and $P_{GMV}$: the proportions of items retrieved by the EBR model in the set of exposed/clicked/purchased items and the proportion in total GMV. As Table 3 shows, MOPPR achieves 0.96% #Transactions and 1.29% GMV improvements on the set of organic search items (distinct from sponsored items and content results). Considering that hundreds of millions of transactions occur on mobile Taobao search every day, these improvements are undoubtedly valuable. MOPPR outperforms MGDSPR with 0.87% $P_{good}$ and 0.13% $P_{h\_good}$ improvements, indicating that MOPPR is better at retrieving relevant items. In addition, based on relevance annotations of exposed items judged by human experts, we find that MOPPR has competitive performances on relevance metrics, much close to the lexical matching method. MOPPR achieves 11.5% $P_{exposure}$, 9.9% $P_{click}$, 6.1% $P_{purchase}$, and 5.7% $P_{GMV}$ absolute improvements, and outperforms all other online retrieval methods on these proportions. All findings reveal that the problems of poor relevance and weak personalization for our EBR model have been addressed to a great extent. With the enhancement of MOPPR, embedding-based retrieval becomes the most important method among all retrieval methods in Taobao search engine. Since the Double-11 shopping festival of 2021, MOPPR has been fully deployed to serve all users of mobile Taobao search, replacing the original MGDSPR.

Table 3: Average improvements of MOPPR over the baseline MGDSPR per day in the 28-day online A/B test, where #Transactions denotes the number of transactions. All improvements are statistically significant examined by paired t-test with $p < 0.01$.

| #Transactions | GMV | $P_{good}$ | $P_{h\_good}$ |
|---|---|---|---|
| +0.96% | +1.29% | +0.87% | +0.13% |
| $P_{exposure}$ | $P_{click}$ | $P_{purchase}$ | $P_{GMV}$ |
| +11.5% | +9.9% | +6.1% | +5.7% |

## 7 ADVANCED TOPICS

In this section, we share our deeper explorations on further improving MOPPR's retrieval performance and introducing the multi-objective optimization into our coarse-grained ranking model.

### 7.1 Hard Mining

*7.1.1 Online Hard Mining.* Following the online hard mining approach proposed by Huang et al. [10], we extend the negatives of a sample for all four objectives by using the positive items of other samples within the same mini-batch. However, we find this approach brings no improvement to MOPPR's offline and online retrieval performances. In our view, this is because most of these negative items in mini-batches of randomly shuffled samples are completely irrelevant to the target query and hence are too easy to used for model training.

*7.1.2 Offline Hard Mining.* We investigate how to effectively mine hard negatives for model learning in an offline way. One idea is to increase the number of under-impressions in the training samples. We increase the number from 10 to 30 and find that this enhances MOPPR with 1.0% improvement on $recall@6000$ and 0.8% improvement on $recall_p@6000$. The second idea is to collect irrelevant items of a certain query from historical logs of impressions and under-impressions and use these items as hard negatives in the training samples with the same query. We find that this approach improves MOPPR on the relevance metric $P_{good}$ with 0.6% improvement.

### 7.2 GMV Maximization Retrieval

Increasing the gross merchandise volume (GMV) is the most concerned goal of e-commerce platforms, because this is directly related to platform revenue. Therefore, we study how to incorporate the GMV maximization objective into MOPPR. We address this problem by using a combined pointwise cross-entropy and multi-objective softmax loss. In the pointwise cross entropy loss, the prediction $\tilde{P}(Purchase = 1|u, q, \mathcal{B}, i)$ for an item $i$ is defined as follows:

$$\tilde{P}(Purchase = 1|u, q, \mathcal{B}, i) = \text{Sigmoid}(\frac{z}{\sigma}) = \frac{1}{1 + \exp(-\frac{z}{\sigma})} \quad (22)$$

where $\sigma$ is a hyperparameter for scaling the inner-product score $z$. We let $\tilde{P} = \tilde{P}(Purchase = 1|u, q, \mathcal{B}, i)$ for convenience. Considering that the average number of purchased items per sample in our data is only 0.04, $\tilde{P}$ has an extremely high probability to be close to zero. Thus, we make an approximate derivation as follows:

$$\frac{z}{\sigma} = \ln(\tilde{P}) - \ln(1 - \tilde{P})$$
$$\approx \ln(\tilde{P}) = \ln(\tilde{P} \cdot Price) - \ln(Price)$$
$$= \ln(\widetilde{GMV}) - \ln(Price) \quad (23)$$

where $Price$ is the selling price of the item $i$ and $\widetilde{GMV} = \tilde{P} \cdot Price$ is the predicted GMV. Therefore, we obtain $\widetilde{GMV}$ as follows:

$$\widetilde{GMV} \approx \exp(\frac{z}{\sigma} + \ln(Price))$$
$$= \exp(\langle \overrightarrow{\phi}(u, q, \mathcal{B}), \overrightarrow{\psi}(i) \rangle / \sigma + \ln(Price))$$
$$= \exp(\langle (\overrightarrow{\phi}(u, q, \mathcal{B}), \sigma), (\overrightarrow{\psi}(i), \ln(Price)) \rangle / \sigma) \quad (24)$$

where $\langle \overrightarrow{\phi}, \overrightarrow{\psi} \rangle$ is the inner-product score function $\mathcal{F}(\overrightarrow{\phi}, \overrightarrow{\psi})$. We generate new user-query and item representation vectors, i.e., $(\overrightarrow{\Phi}(u, q, \mathcal{B}), \sigma)$ and $(\overrightarrow{\Psi}(i), \ln(Price))$, by concatenating $\sigma$ and $\ln(Price)$ into the original user-query and item representations respectively. Therefore, the inner product score between the two new vectors can be used as the score for GMV maximization retrieval.

### 7.3 Cascade Model: From Retrieval to Ranking

We introduce the multi-objective optimization of MOPPR into our coarse-grained ranking model (CGR) shown in Figure 1. In detail, we remove the relevance objective and reserve the purchase, click and exposure objectives in the CGR model. This makes the CGR model focus on optimizing its recall metrics. Since we find that most of retrieved items are of relevant categories to the search query, we utilize randomly sampled negatives with relevant categories to train the model, replacing the original in-batch shared random negatives. Finally, we cascade the EBR model and the CGR model as a new multi-objective cascade model and examine its overall performance. We find this cascade model has significant improvements on offline metrics such as $recall@1000$ and $nDCG@1000$, as well as on online transactions and GMV compared to our previous online solution.

## 8 CONCLUSION

In this paper, we aim to address the poor relevance and weak personalization problems of the embedding-based retrieval model in Taobao search system and propose a multi-objective retrieval model named MOPPR. This model has two main improvements compared to existing EBR models: (1) Sample construction: We construct page view-grained training samples (different from the single-positive training samples for existing EBR models) to restore the entire space of the whole item corpus; (2) Multi-objective optimization: We design four hierarchical optimization objectives, including relevance, exposure, click and purchase. MOPPR learns the relations between different kinds of positive and negative items in a training sample at the same time. In the experiment, MOPPR achieves significant improvements over baselines on both offline and online metrics, showing the effectiveness of our multi-objective solution. We find that it is not applicable to excessively increase the number of randomly sampled negative items for the product retrieval task in the e-commerce search scenario. In the deeper explorations, we enhance MOPPR via hard mining and introduce GMV maximization objective into it. The multi-objective solution of MOPPR can be transferred to downstream ranking models. And the cascade retrieval and ranking model with multiple objectives shows excellent performance in our online search system. In the future, we would like to investigate how to adaptively adjust the weights of multiple optimization objectives for different kinds of samples.